\documentclass[runningheads]{svmult}
\usepackage{makeidx}   
\usepackage{graphicx}  
\usepackage{subeqnar}  
\usepackage{multicol}  
\usepackage{cropmark} 
\usepackage{physprbb}  
\begin{document}
\title*{Antimatter from supersymmetric dark matter}

\author{Fiorenza Donato}
\institute{LAPTH, B.P. 110 Chemin de Bellevue, 74941
Annecy--le--Vieux, France}

\maketitle              

\begin{abstract}
We propose low--energy antideuterons in cosmic rays as a new possible 
signature for indirect detection of supersymmetric dark matter.
 Since the energy spectrum of the antiproton secondary component
 is still spoilt by considerable theoretical uncertainties, 
looking for  low--energy antideuterons seems a plausible alternative.
We apply our calculation to  the AMS experiment, when mounted on the
International Spatial Station. If a few low--energy antideuterons will
be discovered by AMS, this should be seriously taken as a clue for the
existence of relic, massive neutralinos in the dark halo of our
Galaxy. 
\footnote{Report on  work done in collaboration with 
N. Fornengo and P. Salati}
\end{abstract}

\newcommand{\beq}{\begin{equation}}
\newcommand{\eeq}{\end{equation}}
\newcommand{\bea}{\begin{eqnarray}}
\newcommand{\ena}{\end{eqnarray}}
\newcommand{\etal}{{\it et al.}}
\newcommand{\ie}{{\it i.e.}}
\newcommand{\lsim}{\mathrel{\mathop{\kern 0pt \rlap
{\raise.2ex\hbox{$<$}}}
\lower.9ex\hbox{\kern-.190em $\sim$}}}
\newcommand{\gsim}{\mathrel{\mathop{\kern 0pt \rlap
{\raise.2ex\hbox{$>$}}}
\lower.9ex\hbox{\kern-.190em $\sim$}}}
\newcommand{\un}[1]{_{\mbox{\scriptsize #1}}}
\newcommand{\puis}[1]{$^{#1}$}

\newcommand{\app}[3]{Astropart.\ Phys.\ {\bf #1}, #3 (#2)}
\newcommand{\hepex}[1]{{\tt hep-ex/#1}}
\newcommand{\hepph}[1]{{\tt hep-ph/#1}}
\newcommand{\astroph}[1]{{\tt astro-ph/#1}}
\newcommand{\prep}[3]{Phys.\ Rep.\ {\bf #1}, #3 (#2)}
\newcommand{\plb}[3]{Phys.\ Lett.\ B\ {\bf #1}, #3 (#2)}
\newcommand{\npb}[3]{Nucl.\ Phys.\ B\ {\bf #1}, #3 (#2)}
\newcommand{\cpc}[3]{Comm.\ Phys.\ Comm.\ {\bf #1}, #3 (#2)}
\newcommand{\apj}[3]{Astrophys.\ J.\ {\bf #1}, #3 (#2)}
\newcommand{\aeta}[3]{Astron.\ {\&}\ Astrophys.\ {\bf #1}, #3 (#2)}
\newcommand{\prl}[3]{Phys.\ Rev.\ Lett. {\bf #1}, #3 (#2)}
\newcommand{\prd}[3]{Phys.\ Rev.\ D\ {\bf #1}, #3 (#2)}
\newcommand{\rmp}[3]{Rev.\ Mod.\ Phys.\ {\bf #1}, #3 (#2)}
\newcommand{\rnc}[3]{Riv.\ Nuovo\ Cim.\ {\bf #1}, #3 (#2)}
\newcommand{\zfpc}[3]{Z.\ Phys.\ C\ {\bf #1}, #3 (#2)}
\newcommand{\href}[2]{#1}
\newcommand{\email}[1]{\tt #1}

\newcommand{\DFLUX}
{\mbox{$\rm m^{-2} \; s^{-1} \; sr^{-1} \; GeV^{-1}$}} 
\newcommand{\FLUX}{\mbox{$\rm m^{-2} \; s^{-1} \; sr^{-1}$}}
\newcommand{\pbar}{\mbox{$\bar{\rm p}$}}
\newcommand{\nbar}{\mbox{$\bar{\rm n}$}}
\newcommand{\dbar}{\mbox{$\bar{\rm D}$}}
\newcommand{\kX}{\mbox{$\vec{k_{\rm X}}$}}
\newcommand{\kchi}{\mbox{$\vec{k_{\chi}}$}}
\newcommand{\kpbar}{\mbox{$\vec{k_{\pbar}}$}}
\newcommand{\knbar}{\mbox{$\vec{k_{\nbar}}$}}
\newcommand{\kdbar}{\mbox{$\vec{k_{\dbar}}$}}
\newcommand{\kdif}{\mbox{$\vec{\Delta}$}}
\newcommand{\STOT}{\mbox{$\sigma_{\rm tot}$}}
\newcommand{\STOTPP}{\mbox{$\sigma_{\rm p - p}^{\rm tot}$}}
\newcommand{\mpt}{\mbox{$m_{\rm p}$}}
\newcommand{\mnt}{\mbox{$m_{\rm n}$}}
\newcommand{\md}{\mbox{$m_{\dbar}$}}
\newcommand{\pc}{\mbox{$P_{\rm coal}$}}
\newcommand{\ecm}{\mbox{$\sqrt{s}$}}
\newcommand{\np}{\mbox{$n_{\rm p}$}}
\newcommand{\nC}{\mbox{$n_{\chi}$}}
\newcommand{\Ep}{\mbox{$E_{\rm p}$}}
\newcommand{\Epbar}{\mbox{$E_{\pbar}$}}
\newcommand{\Edbar}{\mbox{$E_{\dbar}$}}
\newcommand{\EC}{\mbox{$E_{\chi}$}}
\newcommand{\mC}{\mbox{$m_{\chi}$}}

\section{Introduction.}
\label{sec:introduction}

The dark matter in the Universe and, in particular, in our Galaxy,
could be  mostly made of  WIMPs (Weakly Interacting
Massive Particles). The most appealing candidates to WIMPs are
believed to be the heavy and
neutral species predicted by supersymmetry, called neutralinos. 
The mutual annihilations of these relics in the
halo of our Galaxy would produce an excess in the cosmic radiation
of gamma rays, antiprotons and positrons. In particular, supersymmetric
antiprotons should be abundant at low energy, a region where the flux of
{\pbar} secondaries is a priori negligible. There is quite an excitement trying
to extract from the observations a possible {\pbar} exotic component
which would signal the presence of supersymmetric dark matter
in the Galaxy. Unfortunately, it has been recently realized
\cite{bottino98,bergstrom99,bieber99} that a few processes add up
together to flatten out, at low energy, the spectrum of secondary
antiprotons. Disentangling
an exotic supersymmetric contribution from the conventional
component of spallation antiprotons seems  to be a very
difficult task. 

Antideuterons, \ie, the nuclei of antideuterium, are free from such problems.
 The two antinucleons must be
at rest with respect to each other in order for fusion to take place
successfully. For kinematic reasons, a spallation reaction creates very
few low--energy particles and low-energy secondary antideuterons are
very strongly suppressed.
On the other hand, in neutralino annihilations, antinucleons are
predominantly produced with low energies and  supersymmetric {\dbar}'s 
are manufactured at rest with respect to the Galaxy. This feature is further
enhanced by their subsequent fusion into antideuterons: 
a fairly flat spectrum for supersymmetric antideuterium nuclei is expected.

Below a few GeV/n, secondary antideuterons are quite
suppressed with respect to their supersymmetric partners. That
low--energy suppression is orders of magnitude more effective
for antideuterons than for antiprotons. This makes
cosmic--ray antideuterons a much better probe of supersymmetric
dark matter than antiprotons.

Unfortunately, antideuteron fluxes are quite small with respect
to {\pbar}'s. We will nevertheless show in Sect.~\ref{sec:conclusion}
that a significant portion of the supersymmetric parameter space may
be explored by measuring the cosmic--ray {\dbar} flux at low energy, 
in particular by an AMS/ISS caliber experiment.

\section{Production and propagation of antideuterons.}
\label{sec:coalescence}

 The processes at stake are both the spallation of a cosmic--ray 
high--energy proton on an  atom at rest in the interstellar medium
(secondary component) and the annihilation of a neutralino pair
in the galactic dark halo (primary component).
For each of the processes under concern, the differential probability
for the production of an antiproton or an antineutron may be derived.
The calculation of the probability for the formation of an antideuteron
can proceed in two steps. We first need to estimate the probability for
the creation of an antiproton--antineutron pair. Then, those antinucleons
merge together to yield an antinucleus of deuterium.
A fully description of the processes at stake is carried in 
Ref.\ cite{noi}.

Here we only report the final formula which we obtained for the 
 differential multiplicity of an antideuteron when originated from a 
neutralino pair annihilation:
\beq
{\displaystyle \frac{dN_{\dbar}}{d \Edbar}}\; = \;
\left( {\displaystyle \frac{4 \, \pc^{3}}{3 \, k_{\dbar}}} \right)
\,\left( {\displaystyle \frac{\md}{m_{\pbar} \, m_{\nbar}}} \right)
\;
{\displaystyle \sum_{\rm F , h}} \, B_{\rm \chi h}^{\rm (F)} \,
\left\{
{\displaystyle \frac{dN_{\pbar}^{\rm h}}{d \Epbar}}
\left( \Epbar = \Edbar / 2 \right)
\right\}^{2} \;\; .
\label{dNdbar_on_dEdbar_susy}
\eeq

It may be expressed as a sum, extending over the various quarks and
gluons h as well as over the different annihilation channels F, of the
square of the antiproton differential multiplicity. That sum is weighted
by the relevant branching ratios. The antineutron and antiproton differential
distributions have been assumed to be identical. Here ${\md},
m_{\pbar}, m_{\nbar}$ are, respectively,  the antideuteron, the antiproton
and the antineutron mass, while $k_{\dbar}$ is the antideuteron
momentum. $\pc$ is a critical value for the two--body reduced system
momentum and is linked to the fusion model.

\begin{figure*}[htb!]
\centerline{
\resizebox{0.7\textwidth}{!}
{\includegraphics*[1.5cm,6.5cm][18.5cm,23.cm]
{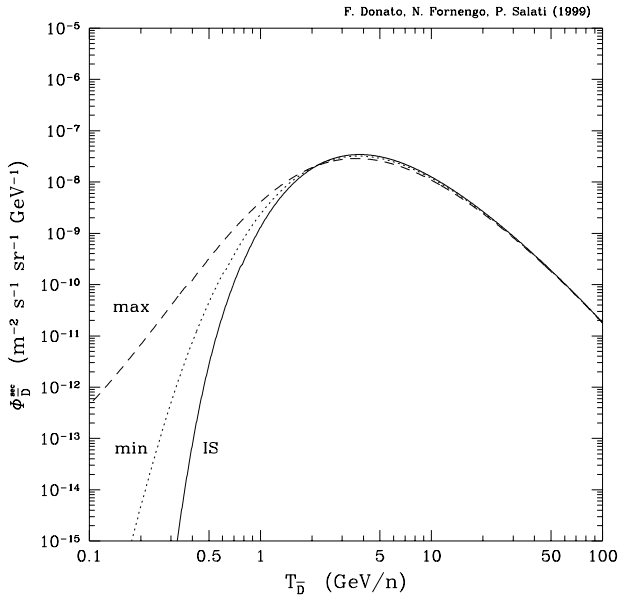}}
}
\caption{Secondary antideuteron spectrum: interstellar (solid curve),
at solar maximum (dashed line) and solar minimum (dotted line).
}
\label{fig:dbar_sec_solmod}
\end{figure*}

The propagation of cosmic--rays inside the Galaxy is
strongly affected by their scattering on the irregularities of magnetic fields,
leading to a diffusive transport. 
Our Galaxy can be reasonably well modelled by a thin disk of atomic and
molecular hydrogen, with radius $R \sim 20$ kpc and thickness $\sim$
200 pc. This gaseous ridge is sandwiched between two diffusion regions
which act as confinement domains as a result of the presence of irregular 
magnetic fields.  They extend vertically up to $\sim 3$ kpc apart from the
central disk. We address to Refs. \cite{noi,bottino98} for a 
careful description of the diffusion model we employed for the
propagation of both the  primary and secondary component. 
 
In Fig.~\ref{fig:dbar_sec_solmod}, the interstellar {\dbar}
 spectrum (solid curve) 
has been modulated at solar maximum (dashed line) and solar minimum
(dotted line). We have applied the forced field approximation \cite{perko}
to estimate the effect of the solar wind on the cosmic--ray energies
and fluxes. We can see that the antideuteron spectrum sharply 
drops below a few GeV/n.
For kinematical reasons, the production of antinucleons at rest with
respect to the Galaxy is extremely unprobable. The manufacture of a
low--energy antideuteron is even more unprobable. It actually requires
the creation of both an antiproton and an antineutron at rest. The momenta
need to be aligned in order for fusion to succesfully take place. Low--energy
antideuterons produced as secondaries in the collisions of high--energy
cosmic--rays with the interstellar material are therefore extremely scarce,
with a completely depleted energy spectrum below $\sim$ 1 GeV/n.
The spallation background is negligible
in the region where supersymmetric {\dbar}'s are expected to be
most abundant. This feature makes the detection of low--energy
antideuterons an interesting signature of the presence of supersymmetric
relics in the Galaxy. 

\begin{figure*}[h!]
\centerline{
{\resizebox{0.5\textwidth}{!}
{\includegraphics*[1.5cm,6.5cm][18.5cm,23.cm]
{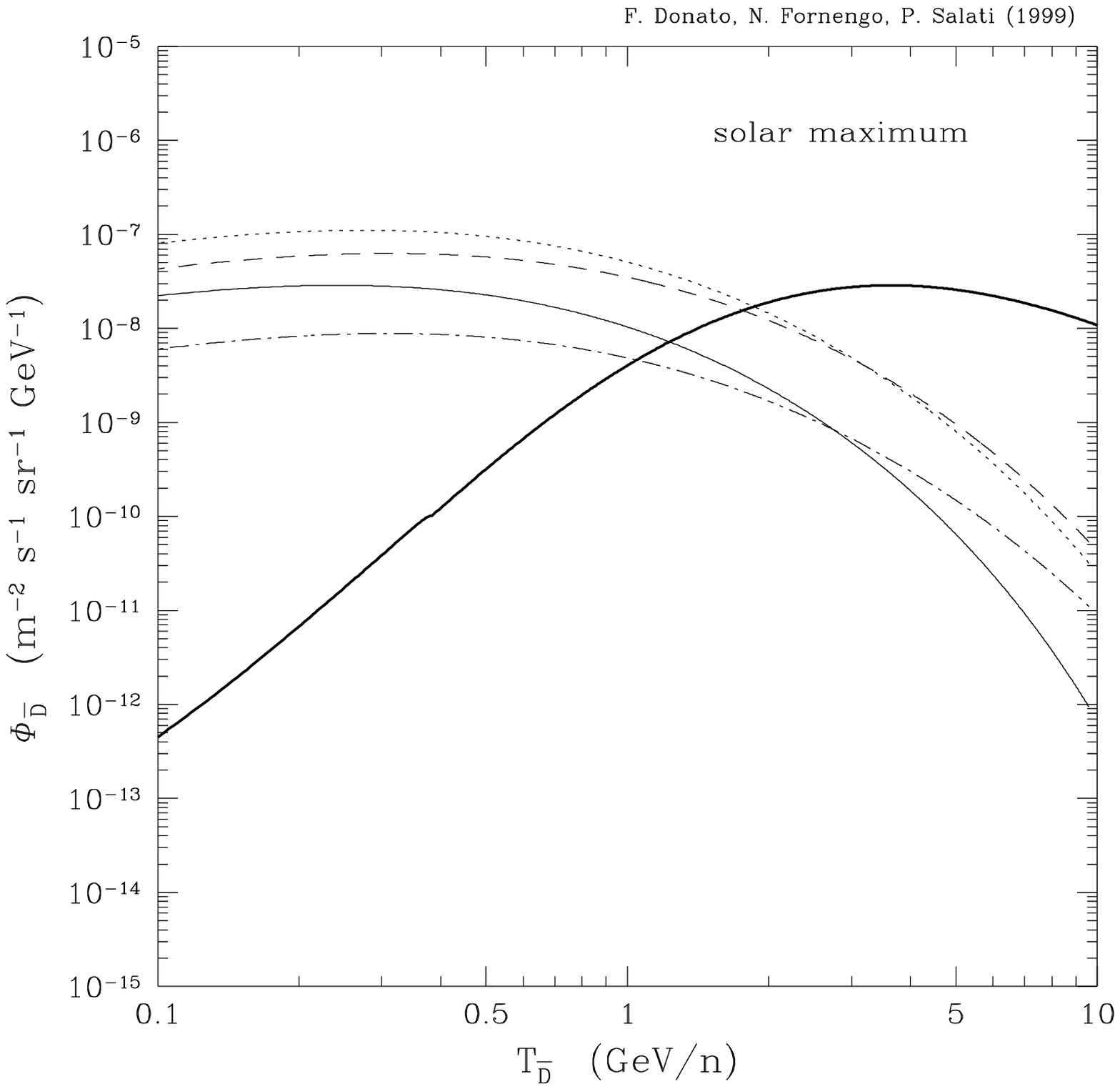}}}
{\resizebox{0.5\textwidth}{!}
{\includegraphics*[1.5cm,6.5cm][18.5cm,23.cm]
{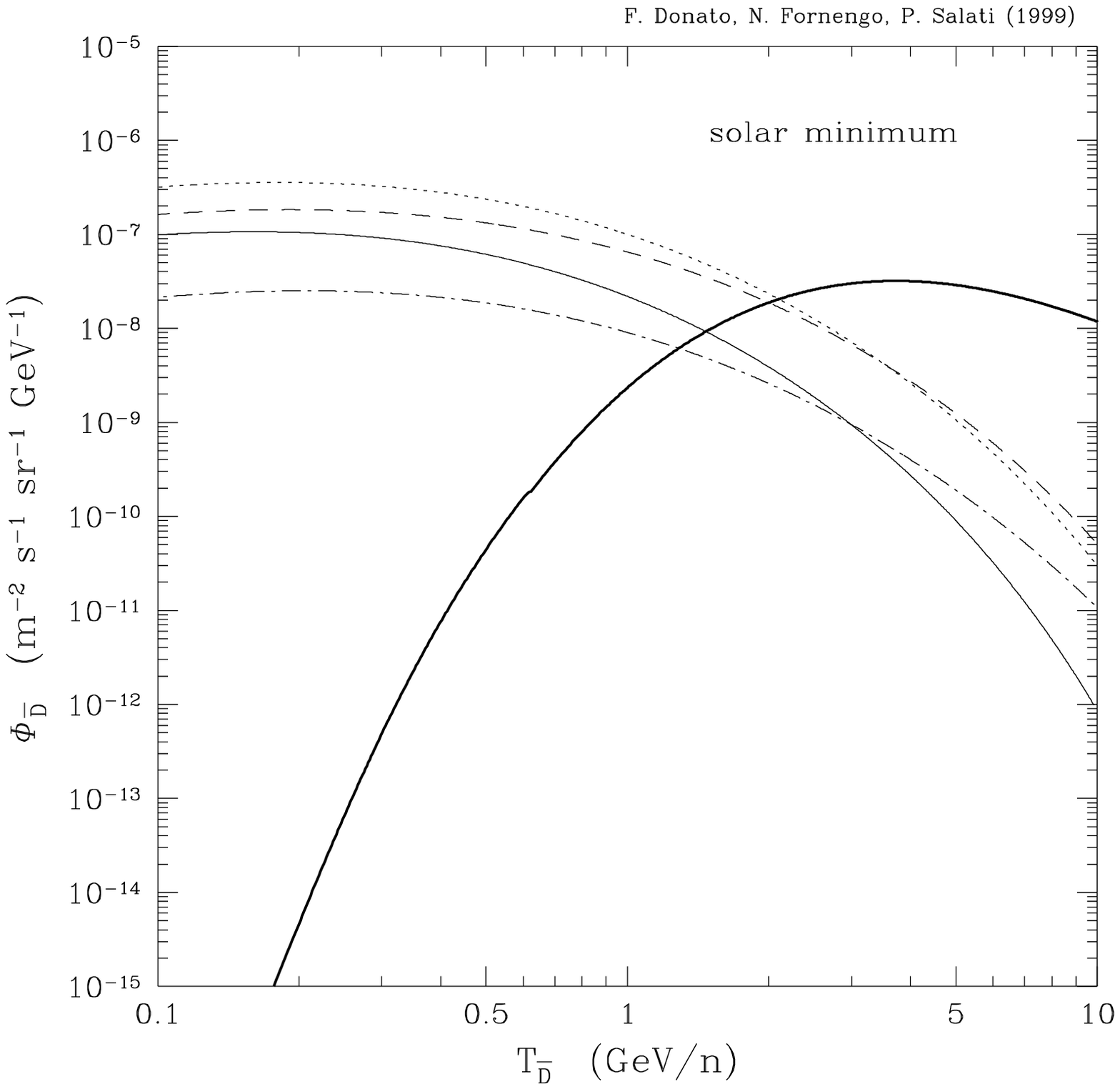}}}
}
\caption{
The solar modulated  flux of secondary antideuterons (heavier solid
curve) compared with the four cases (see text) of antideuterons flux
from from supersymmetric origin.}
\label{fig:dbar_prim_solmod}
\end{figure*}

To this aim, we have calculated the flux
 corresponding to four configurations (see Table 1 of Ref. \cite{noi}
 for details), aimed to illustrate the richness of the supersymmetric
parameter space. 
As a theoretical framework, we use the Minimal Supersymmetric extension of the
Standard Model (MSSM) (see \cite{noi} and references quoted therein)
 which conveniently  describes the
supersymmetric phenomenology at the electroweak scale, without too strong
theoretical assumptions.
We present our results in Fig.\ref{fig:dbar_prim_solmod}.

We can see from this figure that the 
primary fluxes flatten at low energy where they reach a maximum.
As the secondary {\dbar} background vanishes, the supersymmetric
signal is the largest. Neutralino annihilations actually take place at rest
in the galactic frame. The fragmentation and subsequent hadronization of
the jets at stake tend to favour the production of low--energy species.
Therefore, the spectrum of supersymmetric antiprotons -- and antineutrons
-- is fairly flat below $\sim$ 1 GeV. For the same reasons, the coalescence of
the primary antideuterons produced in neutralino annihilations
predominantly takes place with the two antinucleons at rest, hence a flat
spectrum at low energy.

\begin{figure*}[h!]
\centerline{
\resizebox{0.7\textwidth}{!}
{\includegraphics*[1.5cm,6.5cm][18.5cm,23.cm]
{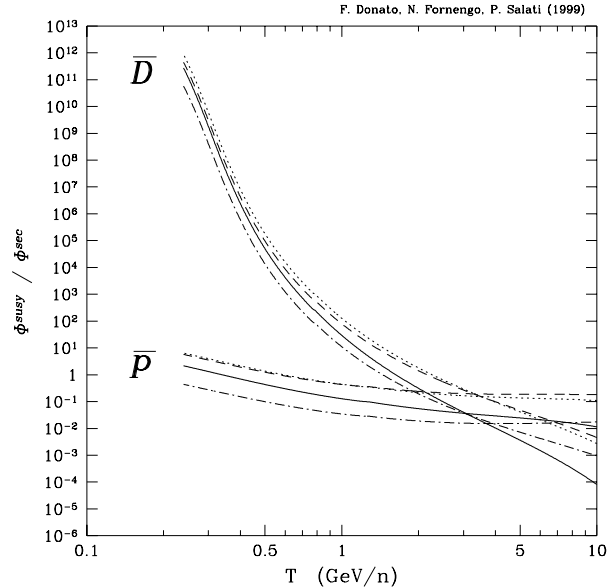}
}}
\caption{
The supersymmetric--to--secondary IS flux ratio for antiprotons
(lower curves) and antideuterons (upper curves) is presented
as a function of the kinetic energy per nucleon. The supersymmetric
configurations are those described in the text.
}
\label{fig:susy_over_sec}
\end{figure*}

In Fig.~\ref{fig:susy_over_sec}, the supersymmetric--to--spallation 
interstellar flux ratios
for antiprotons (lower curves) and antideuterons (upper curves) are presented
as a function of the kinetic energy per nucleon, for the same
configurations as before. In the case of antiprotons, the 
primary--to--secondary ratio is much smaller than for antideuterons.
 The supersymmetric
antiproton signal is swamped in the flux of the secondaries. This is
not the case for
antideuterons. At low energies, their supersymmetric flux is several orders of
magnitude above background. Antideuterons appear therefore as a much cleaner
probe of the presence of supersymmetric relics in the galactic halo
than antiprotons.
The price to pay however is a much smaller flux.
 It is therefore crucial to ascertain which portion of the
supersymmetric configurations will be accessible to future experiments
through the detection of low--energy cosmic--ray antideuterons.

\section{Discussion and conclusions.}
\label{sec:conclusion}

In order to be specific, we have estimated the amount of antideuterons
which may be collected by the AMS experiment \cite{ams}, once it is 
on board ISS.
The future space station is scheduled to orbit at 400 km above the sea level,
with an inclination of $\alpha = 52^{\circ}$ with respect to the Earth
equator. A revolution takes about 1.5 hours so that ISS should fly over
the same spot every day. The AMS detector may be pictured as a 
cylindrical magnetic field with diameter $D = 110$ cm. At any time, its axis 
points towards the local vertical direction. 

Since the terrestrial  magnetic field
prevents particles from penetrating downwards, at any
given geomagnetic latitude $\varrho$ there exists a rigidity cut-off
${\cal R}_{\rm min}$ below which the cosmic--ray flux is suppressed.
The value of the effective solid angle
 through which the arriving particles are potentially detectable
 depends on the cosmic--ray rigidity $p$ as well as on the precise 
location of the detector along the orbit (see \cite{noi} for any detail).

The net number of cosmic--ray species which AMS may collect on
board ISS is actually a convolution of the detector acceptance
$\aleph$ with
the relevant differential flux at Earth. For antideuterons, this leads to
\beq
N_{\dbar} \; = \;
{\displaystyle \int} \aleph \left( p_{\dbar}^{\oplus} \right) \,
\Phi_{\dbar}^{\oplus} \, dT_{\dbar}^{\oplus} \;\; ,
\label{dbar_convolution}
\eeq
where the integral runs on the {\dbar} modulated energy $T_{\dbar}^{\oplus}$.

Integrating the secondary flux previously discussed 
leads respectively to a total of 12.3 and 13.4 antideuterons, depending on
whether the solar cycle is at maximum or minimum. These spallation
{\dbar}'s are mostly expected at high energies. As is clear from
Fig.~\ref{fig:dbar_prim_solmod}, the secondary
flux drops below the supersymmetric signal below a few GeV/n. The transition
typically takes place for an interstellar energy of 3 GeV/n. Below that value,
the secondary antideuteron signal amounts to a total of only 0.6 (solar
maximum) and 0.8 (solar minimum) nuclei.
Most of the supersymmetric signal is therefore concentrated in a low--energy
band extending from the AMS threshold of 100 MeV/n up to an upper
bound of 3 GeV/n in interstellar space. 

\begin{figure*}[h!]
\centerline{
\resizebox{0.7\textwidth}{!}
{\includegraphics*[1.5cm,6.5cm][18.5cm,23.cm]
{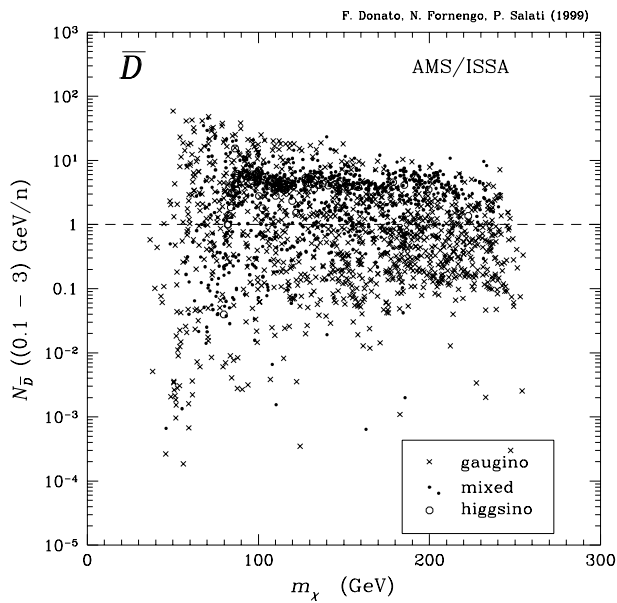}}
}
\caption{
The number $N_{\dbar}$ of antideuterons which AMS
on board ISS can collect between  0.1 up to 3 GeV/n,
 plotted as a function of the neutralino mass $\mC$.}
\label{fig:ndbar_versus_mchi}
\end{figure*}

For each supersymmetric configuration, the {\dbar} flux has
been integrated over that low--energy range. The resulting yield $N_{\dbar}$
which AMS may collect on board ISS is presented as a function of the
neutralino mass $\mC$ in the scatter plot of Fig.~\ref{fig:ndbar_versus_mchi}.
During the AMS mission, the solar cycle will be at maximum.
 A significant portion of the parameter
space is associated to a signal exceeding one antideuteron -- horizontal
dashed line.

\begin{figure*}[h]
\centerline{
{\resizebox{0.5\textwidth}{!}
{\includegraphics*[1.5cm,6.5cm][18.5cm,23.cm]
{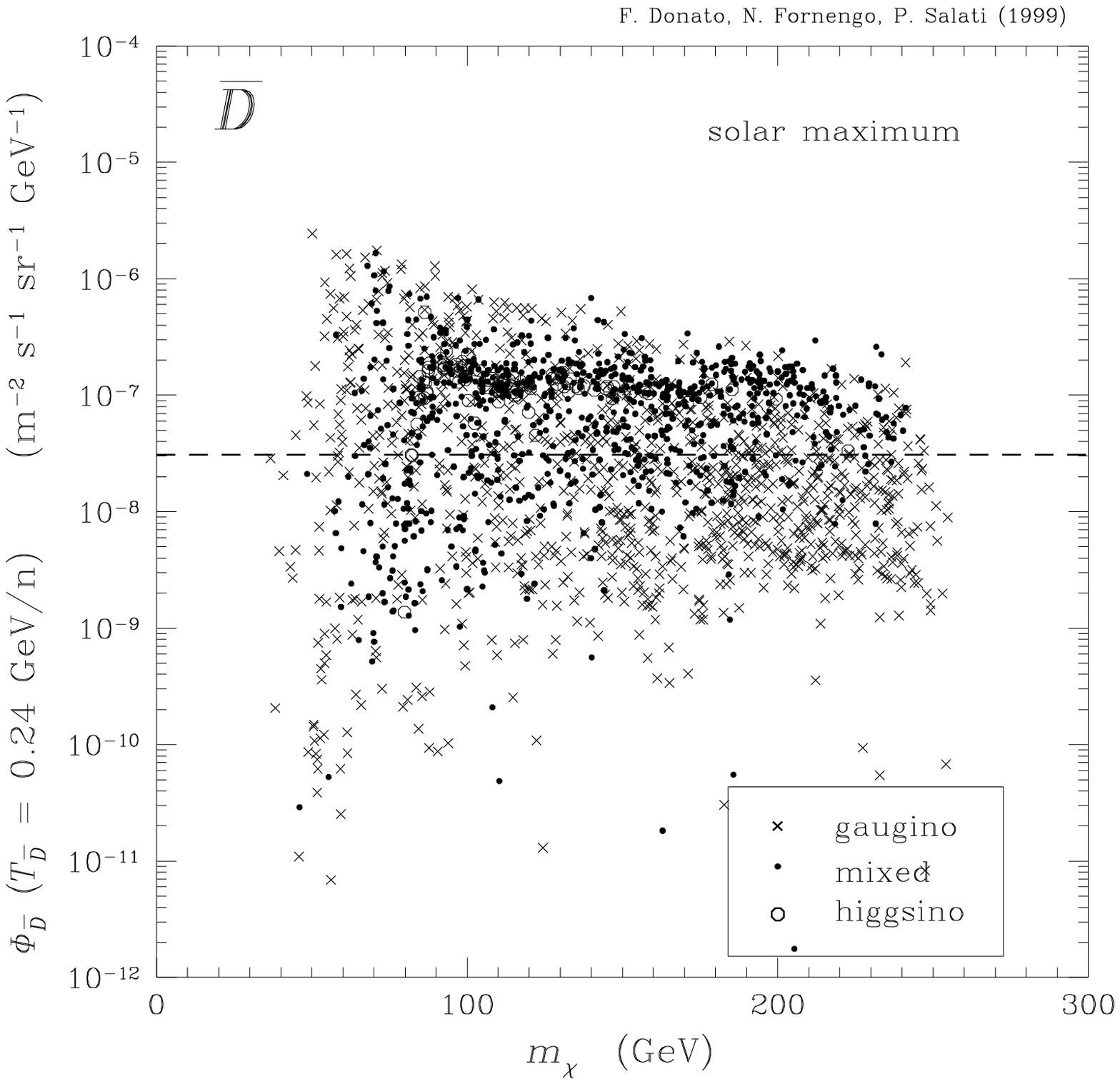}}}
{\resizebox{0.5\textwidth}{!}
{\includegraphics*[1.5cm,6.5cm][18.5cm,23.cm]
{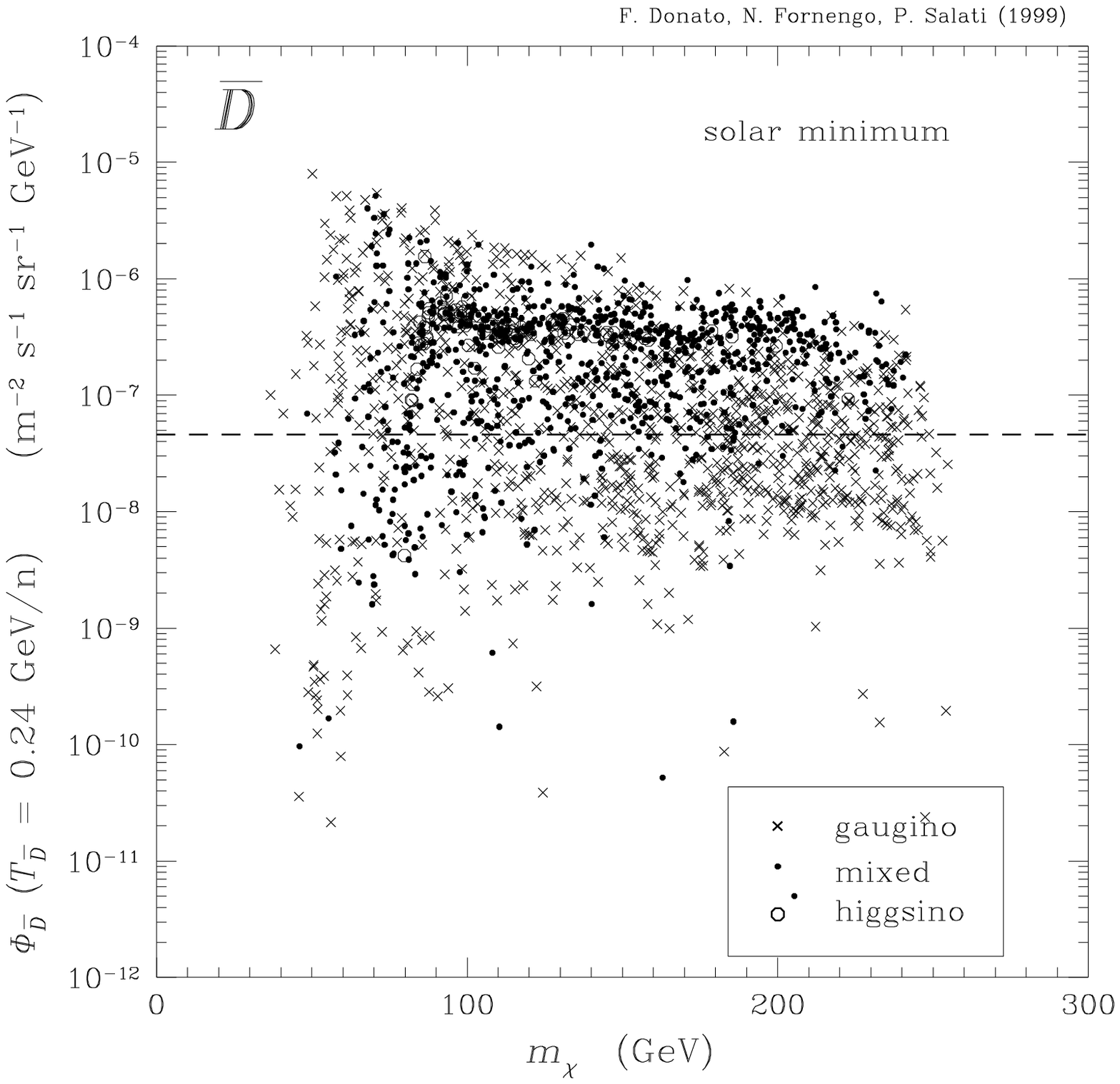}}}
}
\caption{
Scatter plots in the plane $m_{\chi}$--$\Phi_{\dbar}^{\oplus}$. The Earth
antideuteron flux $\Phi_{\dbar}^{\oplus}$ has been computed at solar
maximum ({\rm left}) and minimum ({\rm right}), for a modulated
energy of 0.24 GeV/n. Configurations lying above the horizontal lines
correspond to the detection of at least one antideuteron in the range
of interstellar energies 0.1 -- 3 GeV, by an experiment of the AMS caliber
on board ISS.
}
\label{fig:dbar_scatter_solmod}
\end{figure*}

A complementary information may be derived from 
Fig.~\ref{fig:dbar_scatter_solmod},
where the {\dbar} modulated flux is featured as a function of the neutralino
mass $\mC$. The antideuteron energy at the Earth has been set equal
to 240 MeV/n. The flux $\Phi_{\dbar}^{\oplus}$ is larger at solar
minimum -- when modulation is weaker --than at maximum, the
lower the cosmic--ray energy, the larger that effect.
  The supersymmetric
configurations which an antideuteron search may unravel are
nevertheless more numerous at solar minimum. Between the left and
the right panels, the constellation of representative points is actually
shifted upwards and, relative to the limit of sensitivity, the increase
amounts to a factor $\sim$ 2.
In spite of the low fluxes at stake, the antideuteron channel is sensitive
to a respectable number of supersymmetric configurations.

\begin{figure*}[h!]
\centerline{
\resizebox{0.7\textwidth}{!}
{\includegraphics*[1.5cm,6.5cm][18.5cm,23.cm]
{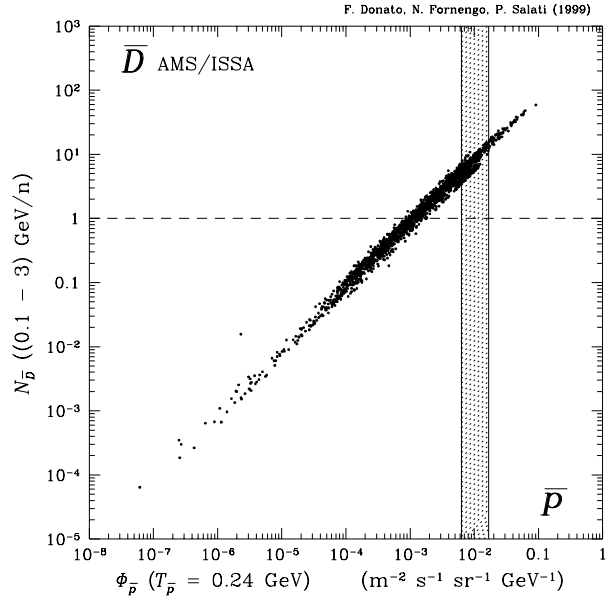}}
}
\caption{
The antideuteron yield $N_{\dbar}$ of
Fig.~\ref{fig:ndbar_versus_mchi}  against the supersymmetric
{\pbar} flux. The antideuteron signal is estimated at solar maximum.
This corresponds to the AMS mission on board the space station. The
{\pbar} flux is derived on the contrary at solar minimum, in the same
conditions as the BESS 95 + 97 flights \cite{bess_2}, whose combined
measurements are indicated by the vertical shaded band for a {\pbar}
energy of 0.24 GeV. The correlation between the antiproton and
antideuteron signals is strong.
}
\label{fig:ndbar_pbar_scatter}
\end{figure*}

 As already  discussed,  it is still a quite difficult task to
ascertain which fraction of the measured antiproton spectrum may
be interpreted as a supersymmetric component. Notice however
that as soon as the secondary {\pbar} flux is reliably estimated,
low--energy antiproton searches will become a more efficient tool.
Meanwhile, we must content ourselves with using observations
as a mere indication of what a supersymmetric component cannot
exceed. The vertical shaded band of Fig.~\ref{fig:ndbar_pbar_scatter}
corresponds actually to the 1--$\sigma$ antiproton flux which the
BESS 95 + 97 experiments \cite{bess_2} have measured at a {\pbar}
kinetic energy of 0.24 GeV.
In Fig.~\ref{fig:ndbar_pbar_scatter}, the supersymmetric
antideuteron yield $N_{\dbar}$ has been derived at solar
maximum. This corresponds to the conditions of the future
AMS mission on board the space station. The antideuteron
yield is plotted as a function of the associated supersymmetric
{\pbar} flux at Earth. The latter is estimated at solar minimum
to conform to the BESS data to which the vertical band refers.
The scatter plot of Fig.~\ref{fig:ndbar_pbar_scatter} illustrates the
strong correlation between the antideuteron and the antiproton
signals, as may be directly guessed from
Eq.~\ref{dNdbar_on_dEdbar_susy}. The horizontal dashed line
indicates the level of sensitivity which AMS/ISS may reach.
Points located above that line but on the left of the shaded vertical
band are supersymmetric configurations that are not yet excluded
by antiproton searches and for which the antideuteron yield is
potentially detectable. The existence of such configurations illustrates
the relevance of an antideuteron search at low energies, 
 appearing  as a plausible alternative, worth
being explored. A dozen spallation antideuterons should be
detected by the future AMS experiment on board ISS
above a few GeV/n. For energies less than $\sim$ 3 GeV/n,
the {\dbar} spallation component becomes negligible and may
be supplanted by a potential supersymmetric signal.
 The discovery of a few low--energy antideuterons
should be taken seriously as a clue for the existence of
massive neutralinos in the dark halo of our Galaxy.

\end{document}